\documentclass[aps,prl,reprint,superscriptaddress]{revtex4-1}

\usepackage{graphicx}
\usepackage{fancyhdr}
\usepackage{dcolumn}
\usepackage{bm}
\usepackage{hyperref}
\usepackage[ margin=0.75in,
]{geometry}

\graphicspath{{./figures/}}

\usepackage[usenames]{color}

\definecolor{OliveGreen}{rgb}{0,0.6,0}

\DeclareMathAlphabet{\mathpzc}{OT1}{pzc}{m}{it}

\newcommand{\CdSe}[1]{\textrm{CdSe$_{\text{(#1)}}$}}

\begin{document}

	\title{Size-dependent lattice dynamics of atomically precise cadmium selenide quantum dots}

	\author{Chenyang Shi}
	\email{cs3000@columbia.edu}
	\affiliation{Department of Applied Physics and Applied Mathematics, Columbia University, New York, NY 10027, USA}
	
    \author{Alexander N. Beecher}
	\affiliation{Department of Chemistry, Columbia University, New York, NY 10027, USA}

	\author{Yan Li}
	\affiliation{American Physical Society, 1 Research Road, Ridge, New York 11961, United States}
	
	\author{Jonathan S. Owen}
	\affiliation{Department of Chemistry, Columbia University, New York, NY 10027, USA}

	\author{Bogdan M. Leu}
    \affiliation{Advanced Photon Source, Argonne National Laboratory, Argonne, IL, 60439, USA}
    \affiliation{Department of Physics, Miami University, Oxford, Ohio 45056, USA}

	\author{Ayman H. Said}
	\author{Michael Y. Hu}
	\affiliation{Advanced Photon Source, Argonne National Laboratory, Argonne, IL, 60439, USA}

	\author{Simon J. L. Billinge}
	\email{sb2896@columbia.edu}
	\affiliation{Department of Applied Physics and Applied Mathematics, Columbia University, New York, NY 10027, USA}
	\affiliation{Condensed Matter Physics and Materials Science Department, Brookhaven National Laboratory, Upton, NY 11973, USA}
	
	\date{\today}

	\begin{abstract}
	\end{abstract}

	\maketitle

{\bf
Material properties depend sensitively on the atomic arrangements and atomic bonding, but these are notoriously difficult to measure in nanosized atomic clusters due to the small size of the objects and the challenge of obtaining bulk samples of identical clusters.
Here we have combined the recent ability to make gram quantities of identical semiconductor quantum-dot nanoparticles with the ability to measure lattice dynamics on small sample quantities of hydrogenated materials using high energy resolution inelastic x-ray scattering (HERIX), to measure the size-dependence of the phonon density of states (PDOS) in CdSe quantum dots. The fact that we have atomically precise
structural models for these nanoparticles allows the calculation of the PDOS using Density Functional Theory (DFT), providing both experimental and theoretical confirmations of the important role that the inertia of the surface capping species plays on determining the lattice dynamics.}

Colloidal semiconducting nanocrystals, commonly called quantum dot nanoparticles, have been studied exhaustively over the last thirty years due to their unique optoelectronic properties: size-tunable band-gaps, narrow, highly efficient photoluminescence, and long-term stability. For these reasons, they have started appearing in various products on the market ranging from television displays~\cite{qdtv;n13,yang;np15}
to solid-state lightbulbs~\cite{mangu;pr17} and biological labels~\cite{supra;mrsbull13, medin;nm05}.
However, despite their commercial success, further development has been hindered by the lack of a detailed understanding of fundamental nanoparticle structure-property relationships, an important example of which is the nature of their lattice dynamics and how it is modified from bulk behavior by nanoparticle size~\cite{beech;jacs16}.
Although lattice dynamics has a significant effect on the structural, mechanical and electronic properties of materials, it remains poorly understood for small crystallites because current techniques are best suited for bulk single-crystals. In the case of nanoparticles, this requirement is particularly problematic because their structures are often poorly defined,  making it challenging even to determine the atomic structure~\cite{billi;s07}, a prerequisite to understanding dynamics and properties.

\begin{figure}	
\includegraphics[width=85mm]{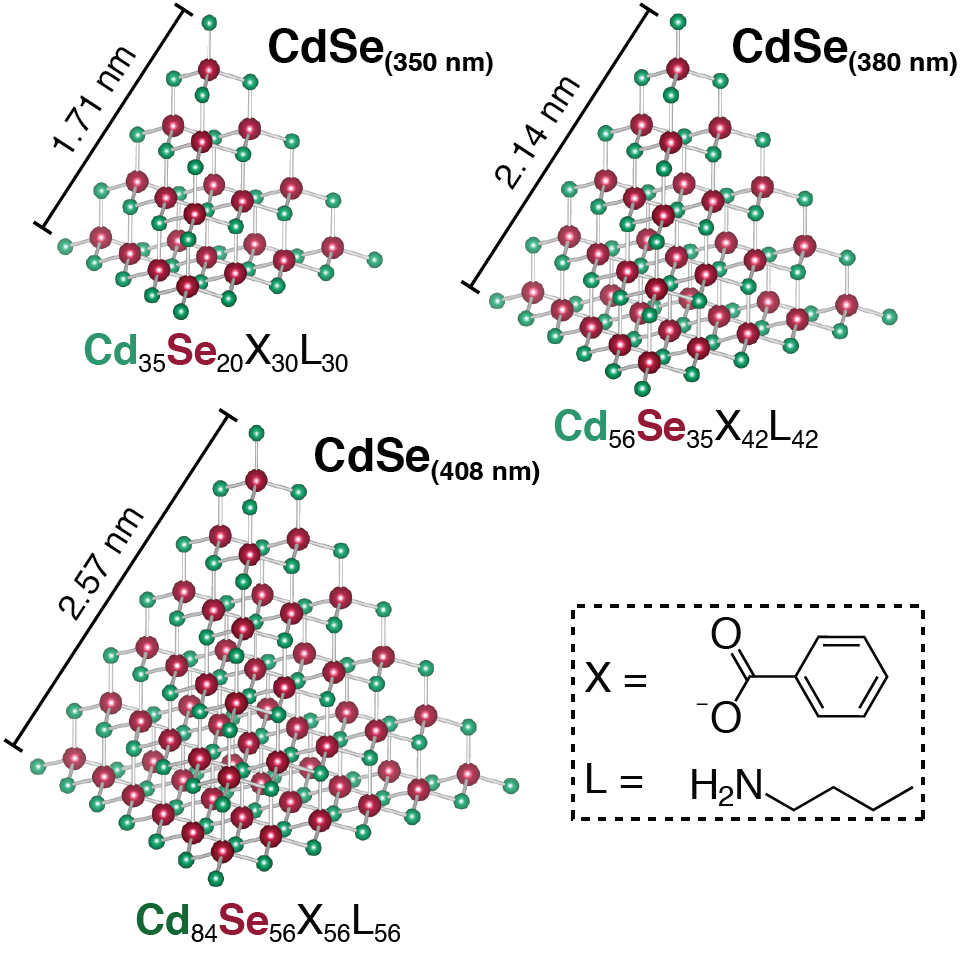}
\caption{Structure representation of the cadmium selenide quantum dots in the current study. Their chemical compositions and surface capping ligands are shown.}
\label{fig:clusters}
\end{figure}

With the development of techniques such as the pair distribution function (PDF) which allows to obtain structure solution from powder x-ray scattering experiments~\cite{juhas;n06}, our ability to measure the structure of nanomaterials has improved. Using the PDF methods, we recently reported the first complete structural solution of a set of atomically-precise cadmium selenide quantum dots~\cite{beech;jacs14}, shown in Fig.~\ref{fig:clusters}.
Unlike even the most monosdisperse nanoparticle samples which generally possess some degree of heterogeneity in size, shape, and/or composition~\cite{chen;nm13}, the class of atomically precise nanoparticles that we prepared have well-defined structures and chemical formulas.
We have isolated three sizes ranging from 1.7-2.6 nm, each of which has a tetrahedral, zincblende cadmium selenide core enriched in cadmium and is passivated by a mixture of benzoate and $n$-butylamine ligands. With precisely known atomic structures, they are ideal candidates for a detailed investigation of lattice dynamics.

Traditionally, inelastic neutron scattering is the tool of choice for studying  phonons in materials since the energy of thermal neutrons is on the order of meV, comparable to the phonon excitation energies~\cite{chapl;b;tps10,dove;b;sd03,dove;b;ild93,widul;pb99}. However, neutrons have disadvantages when studying nanomaterials. One of the major concerns is that a  neutron experiment requires large quantities of the sample, typically on the order of several grams, which is sometimes impossible to satisfy for novel nanoscaled materials.
Moreover for nanomaterials synthesized via a wet chemistry route, hydrogen atoms inevitably exist in the sample, producing a large and problematic
incoherent scattering background. Nuclear resonant inelastic x-ray scattering (NRIXS), which uses isotope resonators such as $^{57}$Fe, $^{119}$Sn, $^{151}$Eu to achieve a better scattering cross section, has been used to study the vibrational properties of some larger nanosystems~\cite{cueny;prb12,stank;prl08} but the small number of suitable isotopes drastically narrows the range of materials that may be studied.
On the other hand, high energy resolution inelastic x-ray scattering (HERIX)~\cite{burke;rpp00,sinn;jpcm01,krisc;c;ixs07,baron;jpcs00} may be carried out on small samples of any material, making it suitable for studies of lattice dynamics in nanoparticles.
For  nanoparticle powder samples,   the wave-vector is no longer well defined due to the breaking of the translational symmetry at finite size, therefore it is only possible to obtain the phonon density of states (PDOS) instead of the full phonon dispersion curves. Nevertheless, the PDOS is still highly valuable for understanding material properties because
a wide variety of thermodynamic quantities such as Gibbs free energy, vibrational entropy, and atomic displacement parameters can be calculated within the harmonic approximation through well-established phonon partition equations~\cite{krisc;c;ixs07,dove;b;sd03,dove;b;ild93}

Here we report a detailed study of the lattice dynamics of three samples of atomically precise quantum dots of different sizes using the HERIX spectrometer at beamline ID-30 at the Advanced Photon Source (APS) at Argonne National Laboratory. Furthermore, knowledge of  the precise core structure of the quantum dots allows us to carry out DFT calculations and understand the measured spectra from a first-principles perspective, giving key insights into modifications to the structure and lattice dynamics of these important nanoparticles.

\section*{\textbf{HERIX \lowercase{experiment and procedure to obtain} PDOS}}
The samples we studied are summarized in Fig.~\ref{fig:clusters} and are labeled \CdSe{350 nm}, \CdSe{380 nm}, and \CdSe{408 nm}, after the first electronic transition peak seen in the ultraviolet-visible absorption spectra~\cite{beech;jacs14}. Details of the synthesis and isolation can be found in our previous paper~\cite{beech;jacs14}. The nanoparticle samples were loaded into Kapton capillary tubes ($d=2$~mm) in a nitrogen-filled glove-box to avoid air exposure, mounted on a copper post, and aligned in the beam.
To optimize the signal a sample thickness was used such that $\mu t = 1$, where $\mu$ is the sample average linear absorption coefficient and $t$ the average sample thickness in the beam. In addition, we also performed measurements on a powdered bulk sample of wurtzite cadmium selenide, \CdSe{Bulk}, which was prepared and mounted in the same way.

During the HERIX experiment, eight silicon diced spherical analyzers~\cite{said;jsr11} independently recorded inelastic x-ray scattering (IXS) signals of the sample, which cover a wide $Q$-range from 54.89~nm$^{-1}$ to 70.57~nm$^{-1}$. The total measurement time for each sample was between 11~and 14~hours. The PDOS data reduction follows previous literature~\cite{kohn;hfi00,bosak;prb05}. Briefly, starting from the summed total IXS pattern, the central elastic peak was subtracted by fitting with a pseudo-Voigt function leaving the inelastic scattering part. A double Fourier-transform method was then applied to simultaneously deconvolute the instrument resolution function and eliminate multi-phonon contributions. Finally, the PDOS is calculated as the product of the single phonon intensity and phonon occupancy factor. More details can be found in the Methods section.

\section*{\textbf{U\lowercase{nexpected broadening and blue-shift of} PDOS \lowercase{with decreasing nanoparticle size}}}
We now proceed to describe the results of the PDOS measurements. The PDOS of a material is expected to be significantly modified when the material becomes nanosized
due to finite size effects. For particles sufficiently small, the continuously dispersing phonon bands are expected to separate into discrete states at well-defined energies corresponding to the
eigenfrequencies of the normal modes of the finite particle. This will result in distinctly different properties of nanoparticles compared to the bulk.
It is therefore of the greatest interest to determine the PDOS of nanoparticles as we have done here.

\begin{figure}
\includegraphics[width=85mm]{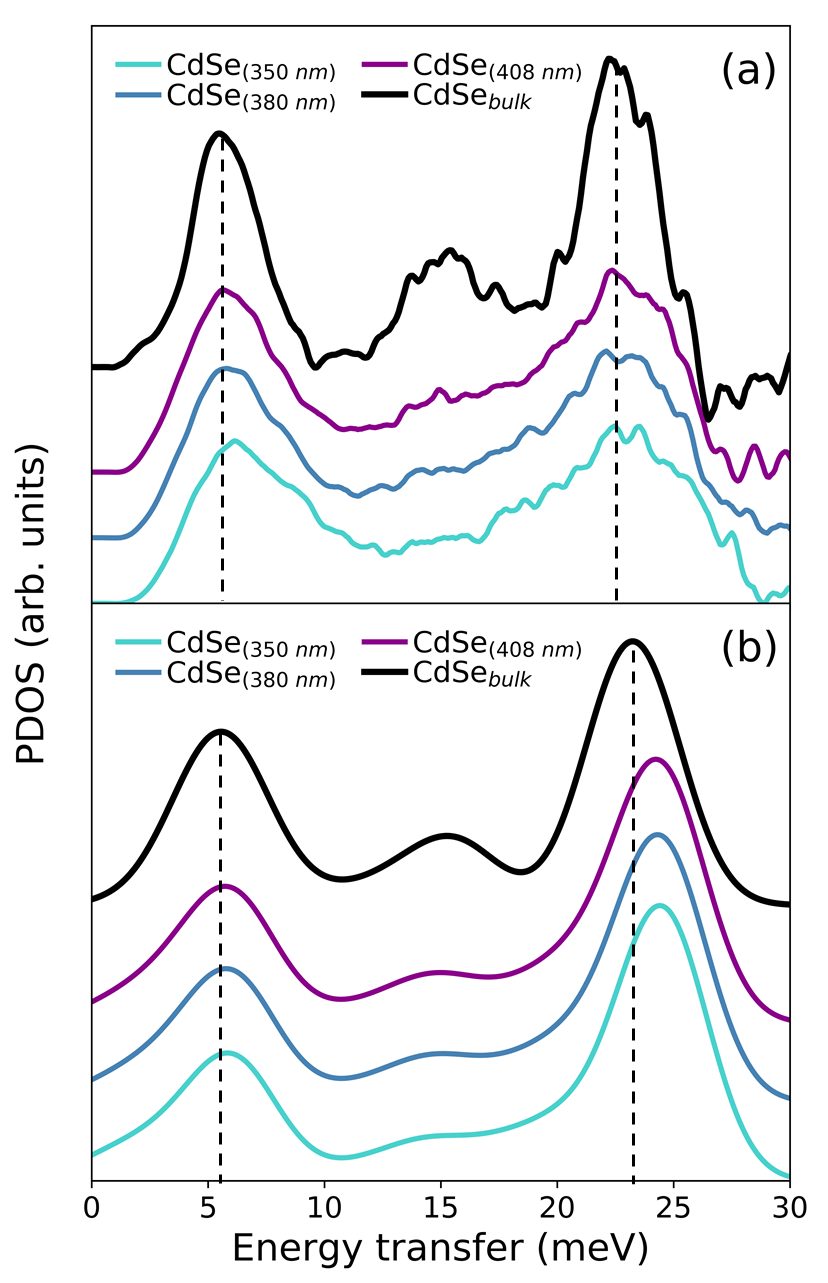}
\caption{ PDOS of the cadmium selenide nanoparticles and  \CdSe{Bulk} from (a) HERIX measurements and (b) DFT calculations. The calculated PDOS curves are normalized with the number of Cd and Se atoms and convoluted with a Gaussian function with a FWHM of 1.709~meV. The dashed lines mark the peak positions at 6~meV and 22~meV from bulk PDOS. Curves are offset vertically for clarity.}
\label{fig:PDOS}
\end{figure}

As a benchmark, we first measured the PDOS from a powdered sample of \CdSe{Bulk}. This is shown as the black line in Fig.~\ref{fig:PDOS}(a). This is in good agreement with the computed PDOS from DFT calculation as shown in Fig.~\ref{fig:PDOS}(b). The calculated PDOS has been convoluted with the resolution function of the measurement with a full width at half maximum (FWHM) of 1.709~meV, which was determined by measuring poly(methyl methacrylate), or PMMA, rod at $Q=10$~nm$^{-1}$. The origin of features in the PDOS can be inferred by a comparison with the computed phonon dispersion curves, as shown in Supplementary Fig. 1.
The low-$\omega$ part of the computed spectrum increases in a power-law fashion due to linearly dispersing acoustic phonons, reaching a broad peak centered around 6~meV in the PDOS where the transverse acoustic (TA) branches bend over as they approach the zone edge. The longitudinal acoustic (LA) modes harden to higher energies before bending over at the zone edge at around 16~meV, resulting in a second broad maximum in the PDOS around this energy. There is a gap between 17 and 21~meV in the phonon spectrum that separates the acoustic branches and the high-$\omega$, weakly dispersing optical  branches, the latter producing a third broad maximum  of intensity centered around 22~meV.
The measured PDOS of \CdSe{Bulk}  (Fig. 2(a)) matches this general shape well, and the measured frequencies are accurately captured in the calculation. Note the energy resolution (1.709~meV) of IXS is not high enough to resolve phonon band gap. In Supplementary Fig 2, this is demonstrated by convoluting  the computed PDOS of \CdSe {Bulk}  with Gaussian functions of various FWHMs.

Next we turn to the measured PDOS of the nanoparticle samples, also shown in Fig.~\ref{fig:PDOS}. A number of qualitative observations can be made. First, they have a shape that generally resembles that of \CdSe{Bulk} with broad peaks around~6 and 22~meV, but the peaks are further broadened out compared to those of the bulk PDOS. Second, in contrast to a well-resolved peak at $\sim$16~meV  in the bulk PDOS, this feature becomes less distinct in the nanoparticle samples. Instead, significant spectral weight now appears in the bulk gap region (17-21~meV).

Despite the overall qualitative agreement between the nanoparticle and bulk PDOSs, a more careful comparison suggests a change in the frequency of some of the phonon modes. In particular, with decreasing particle sizes there is observable spectral weight shifted to higher frequencies for the 6~meV and the 22~meV features. There is no corresponding shoulder on the low-frequency side of the features, and so this results in a slight but definite overall blue-shift of these two features in the PDOS.

\begin{figure}
\includegraphics[width=85mm]{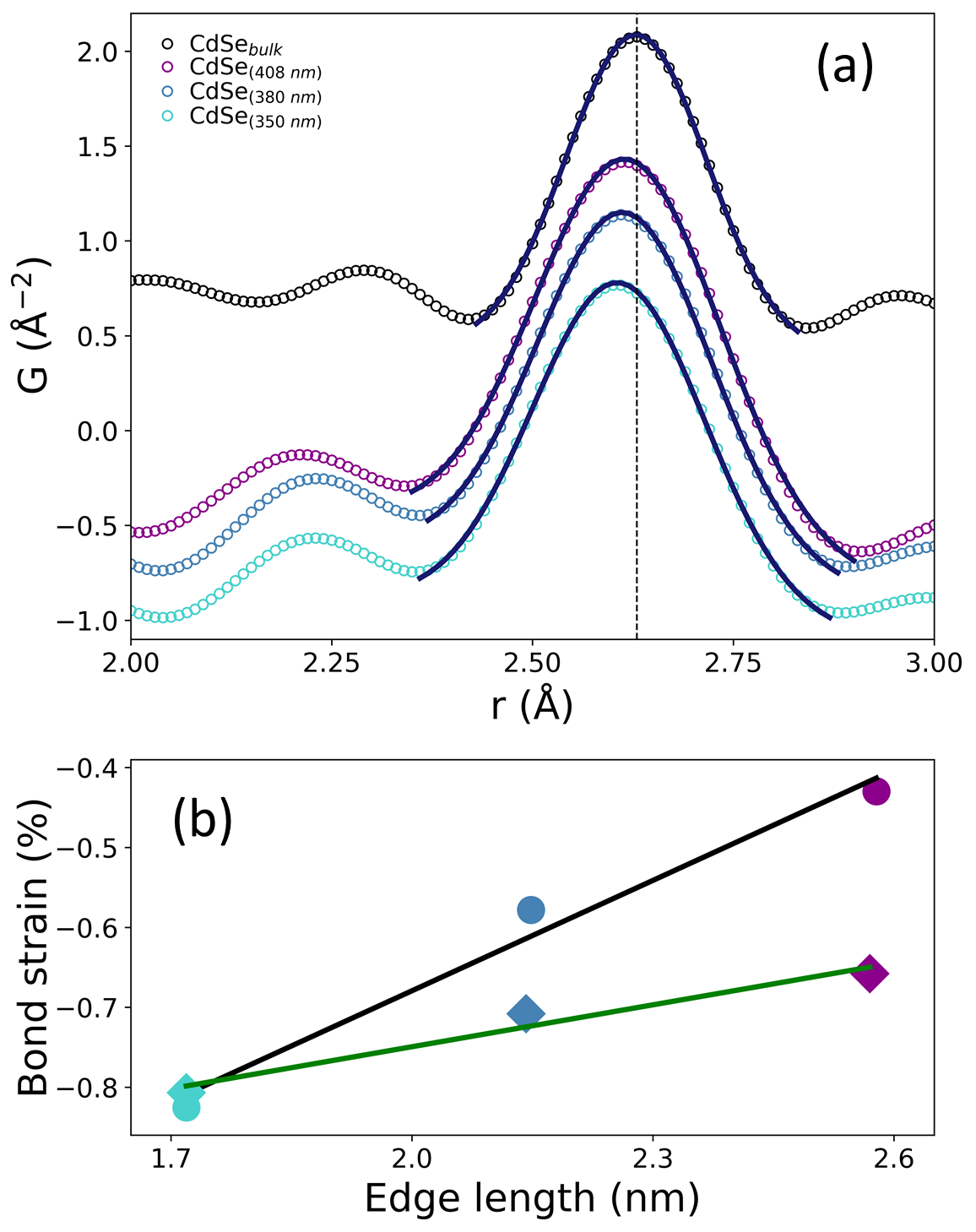}
\caption{(a) Experimental PDFs of the cadmium selenide nanoparticles and \CdSe{Bulk} collected at 20~K. The dashed line marks the peak position of \CdSe {Bulk}. The solid lines are the fits with a Gaussian function. Curves are offset vertically for clarity.  (b) The extracted bond strains as a function of the nanoparticle edge length from PDF (circle) and DFT (diamond).}
\label{fig:PDF}
\end{figure}

Neither the broadening of the PDOS, nor the blue-shift of features in the spectrum, are expected {\it a~priori}.
Finite size effects are expected to result in flatter bands resulting in {\it sharper} features in the PDOS as a result of quantum confinement that leads to localization of the normal modes.
A broadening could come from polydispersity in the sample, with contributions to the signal from different sized nanoparticles.
However, since the samples consist of atomically precise, and therefore identical nanoparticles (one of the features of the current study), we can rule this out.
A broadening could also result from bond relaxations within a single nanoparticle, such that bonds of atoms, for example, at the center or surface of the particle, are of different lengths. There is experimental evidence for such a particle size-dependent Cd-Se peak broadening from PDF studies of small CdSe nanoparticles~\cite{masad;prb07,yang;pccp13,beech;jacs14}.  To confirm this in our samples we show measured PDFs of the current samples and find the same behavior (Fig.~\ref{fig:PDF}).

The blue-shift of features in the PDOS is also interesting. Such an effect has been predicted theoretically~\cite{han;jpcc12}, but never confirmed experimentally.  It is also consistent with another observation from PDF studies of CdSe nanoparticles, i.e.  an overall shortening of the average Cd-Se bond-length with decreasing nanoparticle diameter~\cite{masad;prb07,yang;pccp13,beech;jacs14}. A shorter bond length on average may indicate a stronger bond which oscillates with a higher frequency.

\section*{\textbf{C\lowercase{ombined experiment with} DFT \lowercase{provide insights into structure, bonding and lattice dynamics}}}
The solved atomic structures~\cite{beech;jacs14} of the CdSe nanoparticles under study allow DFT total energy minimization calculations to be carried out to find relaxed structures. It is computationally demanding to simulate vibrational properties of nanoparticles with the full ligand coverage, given the large number of surface ligands and the multitude of different possible ligand arrangements~\cite{vozny;jpcc16} that are not exactly known from the experiments~\cite{beech;jacs14,beech;jacs16}. For simplicity,  the nanoparticle surfaces were passivated with  pseudo-hydrogen atoms (H$^*$) with  fractional charges that fully saturate the dangling bonds of the surface Cd atoms and keep the nanoparticles charge neutral (see Methods section). Although the mass of the terminal group itself does not  affect the optimized DFT geometry of the nanoparticles and their electronic structures, it has a direct influence on the lattice dynamics, i.e.  the phonon frequencies and the weight of the spatial distribution of the eigenmodes, especially those near the surface~\cite{han;jpcc12}. In order to mimic the effects of the real ligands and make direct comparison to the HERIX results, a proper mass of the H$^*$ atoms, $m$,  needs to be chosen. Indeed, when $m$ is simply set to that of a real hydrogen atom ($m=1$~amu), the low-frequency peak near 6~meV in the PDOS  of H$^*$-passivated nanoparticles are found to have a substantial red shift of 1-2 meV (not shown) compared to the bulk while the high-frequency peak near 22~meV is nearly unchanged, in sharp contrast to the experimental results.  On the other hand, when the mass of H$^*$ is chosen to be $m=16$~amu, which is close to the mass of the atoms on the binding ligands (e.g. O in benzoate and N in $n$-butylamine), an excellent agreement with experimental data is established. In particular, the sign of shift relative to the bulk phase changes from negative to positive for both peaks,  with much smaller magnitude ($\sim$0.2~meV) at the low-frequency and much larger at the high-frequency  ($\sim$1 meV).

A higher mass of  $m=32$~amu was also tested to explore the effects of including some of the mass of the rest of the ligand, and was found to have relatively small influence on the PDOS peak positions and magnitude. The influence of different $m$ is also demonstrated in Supplementary Fig. 3 for \CdSe{408 nm}, where the choice of lighter mass ($m=1$~amu)  causes a considerable red shift at both low and high frequencies compared to the case with heavier masses ($m=16$ or 32~amu).  This trend is consistent with that observed for other colloidal nanoparticles with zinc-blende type structures~\cite{han;jpcc12}, although the effects here are more substantial due to the smaller nanoparticles size and larger surface/core ratio considered here.

From the DFT-optimized structures of the nanoparticles and bulk samples, the radial distribution functions (RDF) can be computed and the averaged nearest Cd-Se bond distance can be extracted from the position of the first peak position. These values are used to calculate the bond strains as defined by $[(r_\mathrm{Cd-Se}^\mathrm{Nanoparticle}-r_\mathrm{Cd-Se}^\mathrm{Bulk})/r_\mathrm{Cd-Se}^\mathrm{Bulk}]\times100\%$ where $r_\mathrm{Cd-Se}^\mathrm{Nanoparticle}$ and $r_\mathrm{Cd-Se}^\mathrm{Bulk}$ correspond to the averaged nearest-neighbor Cd-Se bond distances in the nanoparticles and bulk samples, respectively.  A similar calculation was done by using measured PDFs in Fig~\ref{fig:PDF}(a). The experimental and theoretical bond strains are compared in Fig~\ref{fig:PDF}(b), where we see the same trend in both the experiment and the calculation--a reduction of nanoparticle size causes an increased magnitude of compressive strain.
Here the compressive strains for the smallest nanoparticle is $\sim$-0.8\%,  which is slightly smaller than $\sim$-2.0\% for a spherical CdSe nanoparticle with a size of 1.3 nm~\cite{yang;pccp13}.

The PDF signal yields an average bond-strain, but the data are not of high enough resolution to unambiguously determine how particular bonds in the cluster change their length as the clusters get smaller, for example, how the bond-length varies between the core and the surface of the nanoparticles. DFT calculations can give some insight here. With the simple model of H$^*$-termination, a nearly monotonic decrease of the Cd-Se bond from the nanoparticle center to the edge was observed, with the surface bond contracted by roughly 1\% compared to the center, which is in agreement with the observation that the width of the Cd-Se bond-length distribution increases with decreasing particle size in regular quantum dots~\cite{masad;prb07,yang;pccp13}.

In summary, we determined the phonon density of states of recently discovered atomically precise quantum dot particles using synchrotron high energy resolution inelastic x-ray scattering instrumentation.  This is coupled with atomic pair distribution function measurements and first principles density functional theory calculations to provide key insights into the structure, bonding and dynamics of CdSe quantum dot nanoparticles.  Compared to the bulk there is an observable blue-shift in the optical modes which increases with decreasing nanoparticle size.  From theory we show that this blue-shift is very sensitive to the mass of the pseudo-ligand used in the calculation, in fact, the use of hydrogen as the pseudo-ligand results in a red shift of this branch. This implies a strong effect of ligand mass on the dynamical spectrum of the nanoparticle core which might present opportunities for engineering the PDOS spectrum by controlling ligand mass and coverage.

\section*{\textbf{A\lowercase{cknowledgments}}}
Work in the Billinge group was supported by U.S. Department of Energy, Office of Science, Office of Basic Energy Sciences (DOE-BES) under contract No. DE-SC0012704. Work on the nanoparticle synthesis and characterization was supported by the National Science Foundation through contract No. NSF-CHF-1151172. This work made use of the Advanced photon Source which is supported by the U.S. Department of Energy, Office of Science, Office of Basic Energy Sciences through contract No.~DE-AC02-06CH11357. The x-ray PDF measurements were conducted on beamline 28-ID-2 of the National Synchrotron Light Source II, a U.S. Department of Energy (DOE) Office of Science User Facility operated for the DOE Office of Science by Brookhaven National Laboratory under Contract No. DE-SC0012704. Computations made use of resources of the National Energy Research Scientific Computing Center, a DOE Office of Science User Facility supported by the Office of Science of the U.S. Department of Energy under Contract No. DE-AC02-05CH11231.
	
\section*{\textbf{A\lowercase{uthor contributions}}}
S.J.L. and J.S.O conceived the project. A.N.B. synthesized the samples. C.S. and A.N.B. collected X-ray total scattering data. C.S., A.N.B., B.M.L. and A.H.S. measured inelastic X-ray scattering (IXS) data. C.S. analyzed both experimental data. M.Y.H. checked the IXS data analysis. Y.L. performed DFT calculations. C.S., Y.L. and S.J.L. wrote the manuscript, which all the authors discussed.

\section*{\textbf{C\lowercase{ompeting financial interests}}}
The authors declare no competing financial interests.

\section*{\textbf{M\lowercase{ethods}}}
\subsection{HERIX experiments.}

The high energy resolution inelastic x-ray scattering experiments were carried out at room temperature at Sector~30 at Argonne National Laboratory. The \CdSe{Bulk} and nanoparticle powder samples were encapsulated in Kapton capillary tubes. By scanning $x$, $y$ positions along the tubes, the optimum  absorption length was achieved for each sample which maximizes the inelastic scattering signal. The incident x-ray with an energy of 23.7245~keV ($\lambda$=0.5226~\AA) and a beam size of $1.8\times0.4$~mm was unfocused on the samples, and eight silicon spherical analyzers recorded the data simultaneously in a momentum transfer range from 54.89~nm$^{-1}$ to 70.57~nm$^{-1}$ (corresponding to a $2\theta$-range from 26.51$^\circ$~to~33.98$^\circ$). The energy transfer was chosen from -10~meV to +40~meV~\cite{toell;jsr11}. Each sample was measured for 50~minutes and the measurement was repeated at least 11 times before summation. A poly(methyl methacrylate) or PMMA rod was measured at RT at 10~nm$^{-1}$ to determine the resolution function and efficiency of each analyzer. An empty capillary was also measured for background subtraction. The measured IXS spectra were summed into one spectrum using an ``incoherent approximation" where intensity measured by each individual analyzer was divided by the respective analyzer efficiency before summation.

\subsection{PDF experiments.}

Synchrotron x-ray total scattering experiments were conducted at beamline 28-ID-2 at the National Synchrotron Light Source II at Brookhaven National Laboratory. The samples were packed into Kapton capillary tubes and measured at 20~K using a cryostat. The rapid acquisition pair distribution function (RaPDF) technique~\cite{chupa;jac03} was used with an x-ray energy of 67.4194~keV ($\lambda$=0.1839~\AA). A large area 2D Perkin Elmer detector ($2048\times2048$ pixels and $200\times200$ $\mu$m pixel size) was mounted orthogonal to the beam path with a sample-to-detector distance of 212.641~mm for \CdSe{350 nm}, \CdSe{380 nm} and \CdSe{Bulk} while 206.377~mm for \CdSe{408 nm} which were determined by using a nickel standard as a calibrant. The raw 2D data were azimuthally integrated and converted to 1D intensity versus 2${\theta}$ using FIT2D~\cite{hamme;hpr96}. PDFgetX3~\cite{juhas;jac13} was used to correct and normalize the diffraction data and then Fourier transform them to obtain the experimental PDF, \emph{G(r)}, according to $G(r)=2/\pi \int_{Q_{min}}^{Q_{max}}Q[S(Q)-1]\sin Qr\>\mathrm{d} Q$. Here $Q$ is the magnitude of the momentum transfer on scattering and $S(Q)$ is the properly corrected and normalized powder diffraction intensity measured from ${Q_{min}}$ to ${Q_{max}}$~\cite{egami;b;utbp13}.
\\
\subsection{Procedures to obtain PDOS.}

The procedures to obtain PDOS from IXS measurements of powder samples closely follow the literature~\cite{kohn;hfi00,bosak;prb05} where the multi-phonon contribution is eliminated simultaneously with the deconvolution of instrumental function. Starting from the summed IXS spectrum in an ``incoherent approximation", the elastic scattering contribution was carefully subtracted, leaving an inelastic scattering part $I(E)$. Through $\int I(E)dE = I_{0}(1 - f_{LM})$ and $\int I(E)EdE = I_{0}E_{R}$, a scaling factor $I_{0}$ and Lamb-M\"{o}ssbauer factor $f_{LM}$ were obtained. Here $E_{R}$ is the recoil energy of a free nucleus of mass $M$ (a reduced mass is used in our case) and is defined as $E_{R} = \hbar^2k^2/2M$. Next, the Fourier transformation of the normalized resolution function $P(E)$ (i.e. $\int P(E)dE = 1$ and $\int P(E)EdE = 0$) and $I(E)$ were introduced as $Q(\tau)$ and $J_{0}(\tau)$, respectively, where $Q(\tau) = \int \exp(iE\tau)P(E)dE$ and $J_{0}(\tau) = \int \exp(iE\tau)I(E)dE$. A numerical parameter $P_{if}$ (0.1 was used here) is defined as the degree of deconvolution so that $Q_{0}(\tau) = (Q(\tau) + P_{if})/(1 + P_{if})$. After dividing out the resolution function in Fourier space, $M(\tau) = \ln\left(1 + \frac{\int  \exp(iE\tau)I(E)dE}{I_{0}f_{LM}Q_{0}(\tau)}\right)$ was obtained, which corresponds to the Fourier transformation of the single-phonon scattering term. Finally, an inverse Fourier transform of $M(\tau)$ and a multiplication by the phonon occupancy factor, gives PDOS, $g(E)$, according to $g(E) = \frac{E}{E_{R}}[1 - \exp(-E/kT)]\int\frac{1}{2\pi}\exp(-iE\tau)M(\tau)d(\tau)$.

\subsection{Density functional theory calculations.}

DFT  calculations were carried out with the projected augmented wave  method~\cite{bloch;prb94} implemented in the Vienna ab initio simulation package (VASP)~\cite{kress;prb96} package,  with the local density approximations and a kinetic energy cutoff of 300 eV. Three sizes of zinc-blende type CdSe nanoparticles with tetrahedral symmetry ($T_d$) were considered (Cd$_{35}$Se$_{20}$H$^*_{60}$, Cd$_{56}$Se$_{35}$H$^*_{84}$, Cd$_{84}$Se$_{56}$H$^*_{112}$), where  H$^*$ denotes ``pseudo'' hydrogen atoms with a fractional charge of 3/2 in order to passivate the dangling bonds of surface Cd atoms while maintaining the charge neutrality~\cite{deng;prb12}.  The same core formula as  those  shown in Fig.~\ref{fig:clusters} are adopted, but with the ligand shell replaced by H$^*$ atoms and is, therefore, of higher symmetry.  Different atomic mass of H$^*$ atoms ($m$=1, 16, 32) were tested to examine the influence of ligand mass on the vibrational properties of the nanoparticles. Dynamics matrix was computed via the  density-functional perturbation theory~\cite{baron;rmp01}, and the phonon density of states was obtained using the PHONOPY~\cite{togo;prb08} program.

\subsection{Code and Data availability}
The data and computer code that support the findings of this study are available from the corresponding authors upon reasonable request.

	\clearpage
	\newpage
	\onecolumngrid
	\appendix
	\setcounter{figure}{0}
	\renewcommand{\figurename}{Extended Data Figure}

\section{Supplementary information}

\begin{figure*}[!htb]
\includegraphics[width=150mm]{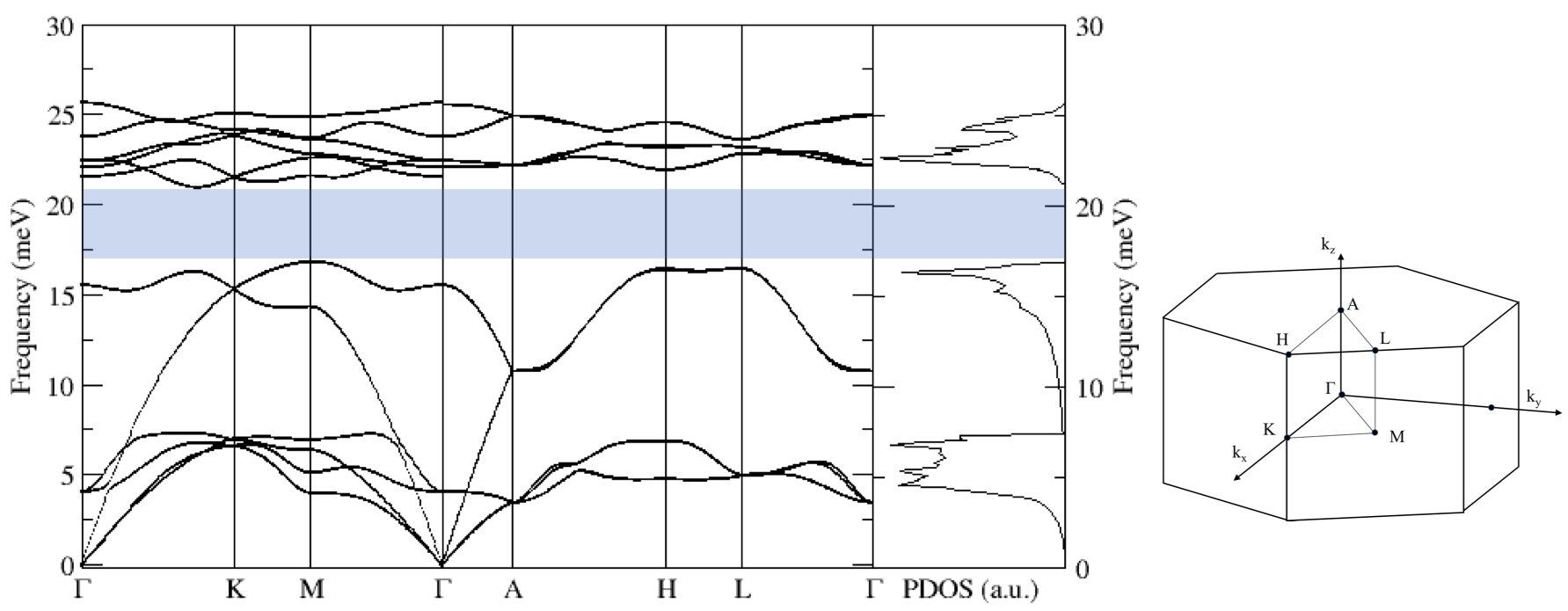}
\caption{Computed phonon dispersion curves along the special directions ($\Gamma$-K-M-$\Gamma$-A-H-L-$\Gamma$) and PDOS of \CdSe{Bulk}. The band gap  between acoustic and optical phonons, between 17 and 21 meV,  is highlighted. Also shown is the Brillouin zone of the hexagonal wurtzite lattice.}\label{fig:dispersion}
\end{figure*}

\begin{figure*}[!htb]
\includegraphics[width=100mm]{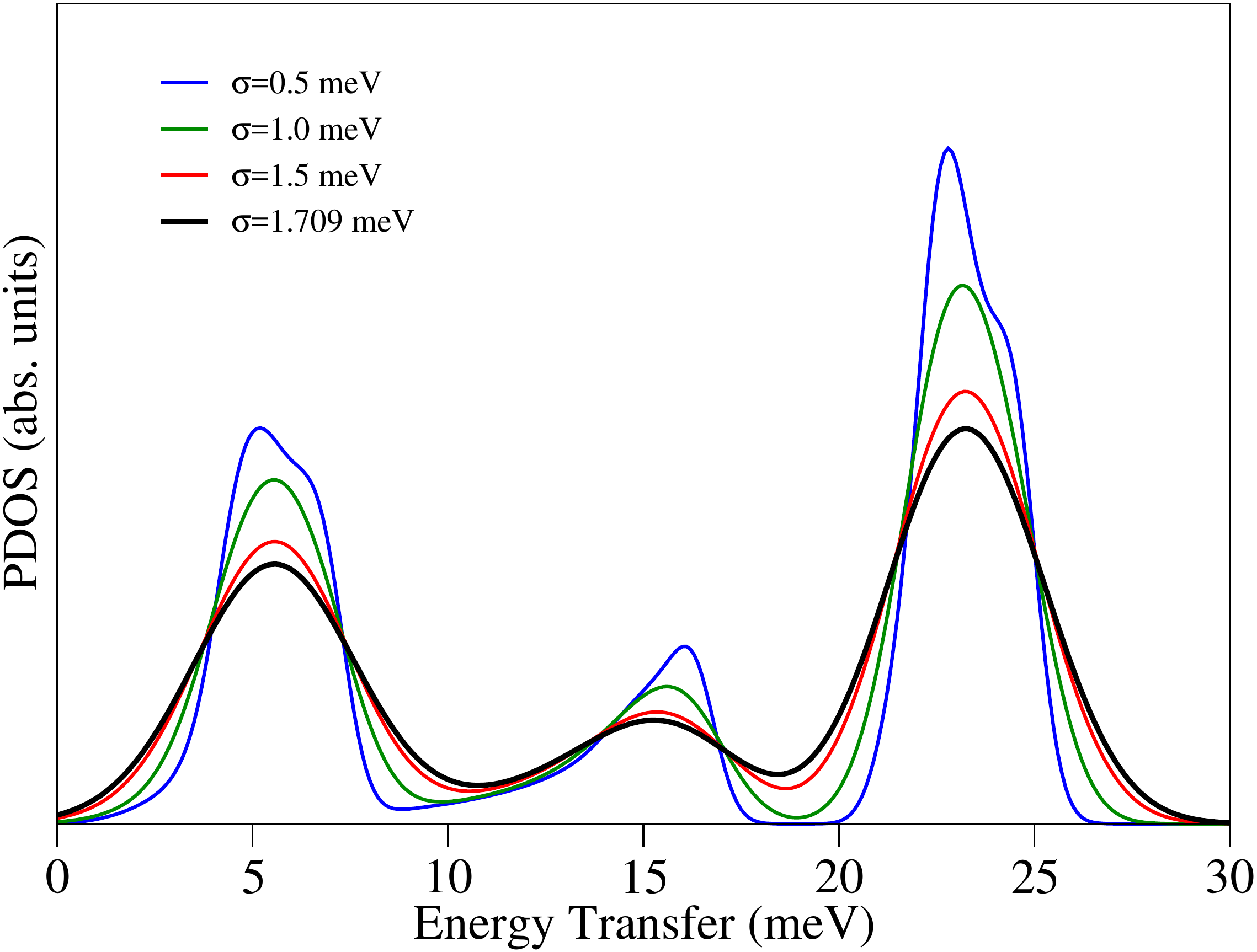}
\caption{Simulated PDOS of \CdSe{Bulk} convoluted with Gaussian functions with various FWHMs.}\label{fig:PDOS_Gauss}
\end{figure*}

\begin{figure*}[!htb]
\includegraphics[width=100mm]{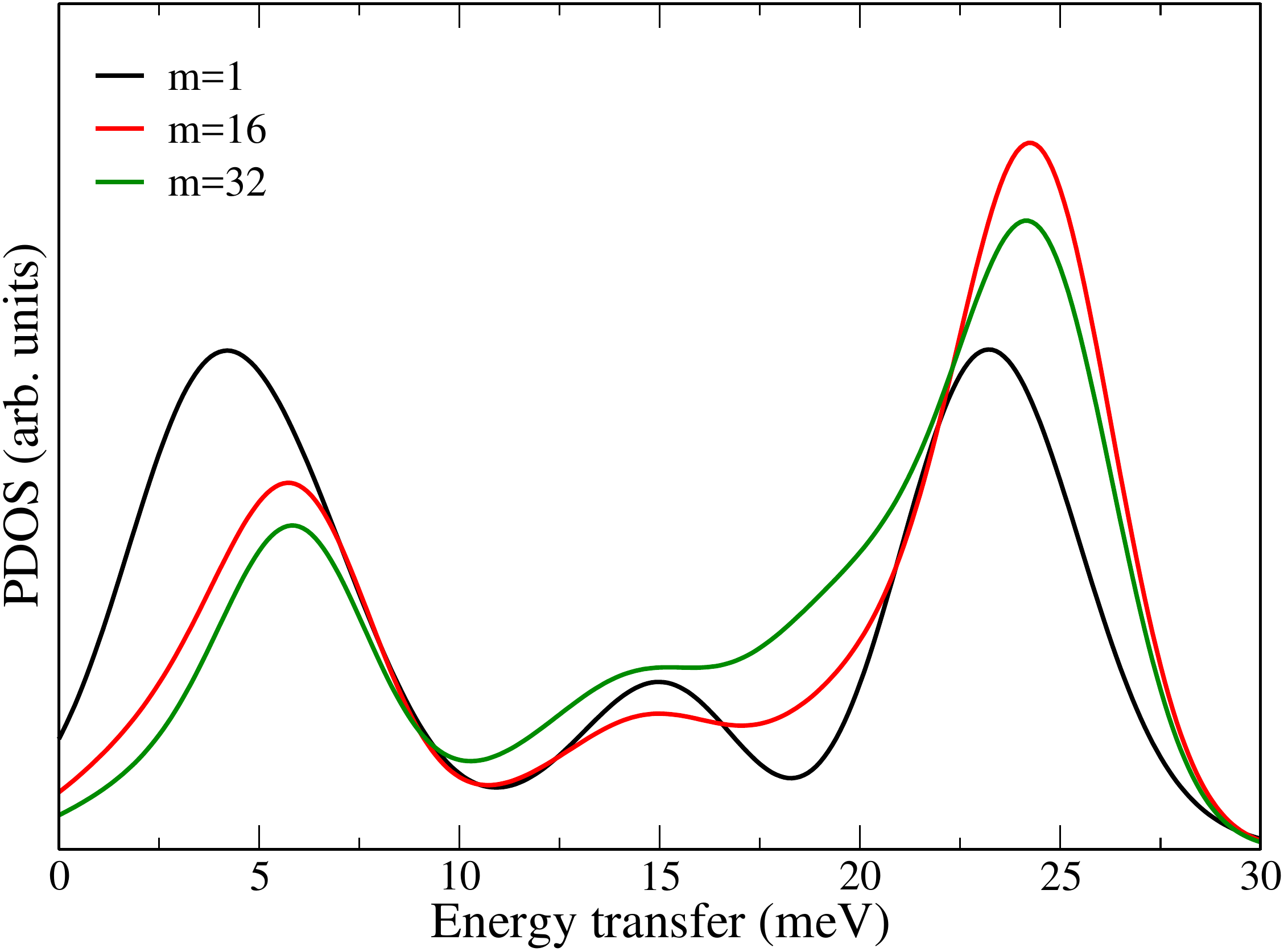}
\caption{PDOS of  Cd$_{84}$Se$_{56}$H$^*_{112}$ computed with different mass for  H$^*$.  The PDOS was convoluted with a Gaussian function with a width of 1.709~meV that came from the instrument resolution.}
\label{fig:PDOS_mass}
\end{figure*}


\begin{thebibliography}{37}

\bibitem{qdtv;n13}
Bourzac, K. Quantum dots go on display.
\newblock{\emph{Nature}} \textbf{493}, 283 (2013).

\bibitem{yang;np15}
Yang, Y. \emph{et al.} High-efficiency light-emitting devices based on quantum dots with tailored nanostructures.
\newblock{\emph{Nat. Photonics}} \textbf{9}, 259-266 (2015).

\bibitem{mangu;pr17}
Mangum, B. D., Landes, T. S., Theobald, B. R. \& Kurtin, J. N. Exploring the bounds of narrow-band quantum dot downconverted LEDs.
\newblock{\emph{Photonics Res.}} \textbf{5}, A13-A22 (2017).

\bibitem{supra;mrsbull13}
Supran, G. J. \emph{et al.} QLEDs for displays and solid-state lighting.
\newblock{\emph{MRS Bull.}} \textbf{38}, 703-711 (2013).

\bibitem{medin;nm05}
Medintz, I. L., Uyeda, H. T., Goldman, E. R. \& Mattoussi H. Quantum dot bioconjugates for imaging, labelling and sensing.
\newblock{\emph{Nature Mater.}} \textbf{4}, 435-446 (2005).

\bibitem{beech;jacs16}
Beecher, A. N., Dziatko, R. A., Steigerwald, M. L., Owen, J. S. \& Crowther, A. C. Transition from molecular vibrations to phonons in atomically precise cadmium selenide quantum dots.
\newblock {\emph{J. Am. Chem. Soc.}} \textbf{138}, 16754-16763 (2016).

\bibitem{billi;s07}
Billinge, S. J. L. \& Levin, I. The problem with determining atomic structure at the nanoscale.
\newblock {\emph{Science}} \textbf{316}, 561-565 (2007).

\bibitem{juhas;n06}
Juh\'as, P., Cherba, D. M., Duxbury, P. M., Punch, W. F. \& Billinge, S. J. L. \emph{Ab initio} determination of solid-state nanostructure.
\newblock {\emph{Nature}} \textbf{440}, 655-658 (2006).

\bibitem{beech;jacs14}
Beecher, A. N. \emph{et al.} Atomic structures and gram scale synthesis of three tetrahedral quantum dots.
\newblock {\emph{J. Am. Chem. Soc.}} \textbf{136}, 10645-10653 (2014).

\bibitem{chen;nm13}
Chen, O. \emph{et al.} Compact high-quality CdSe–CdS core–shell nanocrystals with narrow emission linewidths and suppressed blinking.
\newblock {\emph{Nature Mater.}} \textbf{12}, 445-451 (2013).

\bibitem{chapl;b;tps10}
Chaplot, S. L., Mittal, R. \& Choudhury, N.
\newblock {\emph{Thermodynamic Properties of Solids}}
\newblock (Wiley-VCH Verlag, Weinheim, 2010).

\bibitem{dove;b;sd03}
Dove, M. T.
\newblock {\emph{Structure and Dynamics: An Atomic View of Materials}}
\newblock (Oxford University Press, Oxford, 2003).

\bibitem{dove;b;ild93}
Dove, M. T.
\newblock {\emph{Introduction to Lattice Dynamics}}
\newblock (Cambridge University Press, Cambridge, 1993).

\bibitem{widul;pb99}
Widulle, F. \emph{et al.} The phonon dispersion of wurtzite CdSe.
\newblock {\emph{Physica B}} \textbf{263-264}, 448 (1999).

\bibitem{cueny;prb12}
Cuenya, B. R.\emph{et al.} Size-dependent evolution of the atomic vibrational density of states and thermodynamic properties of isolated Fe nanoparticles.
\newblock {\emph{Phys. Rev. B}} \textbf{86}, 165406 (2012).

\bibitem{stank;prl08}
Stankov, S. \emph{et al.} Vibrational properties of nanograins and interfaces in nanocrystalline materials.
\newblock {\emph{Phys. Rev. Lett.}} \textbf{100}, 233503 (2008).

\bibitem{burke;rpp00}
Burkel, E. Phonon spectroscopy by inelastic x-ray scattering.
\newblock {\emph{Reports Prog. Phys.}} \textbf{63}, 171-232 (2000).

\bibitem{sinn;jpcm01}
Sinn, H. Spectroscopy with meV energy resolution.
\newblock {\emph{J. Phys.: Cond. Matter.}} \textbf{13}, 7525 (2001).

\bibitem{krisc;c;ixs07}
Krisch, M. \& Sette, F.
\newblock {\emph{Light Scattering in Solids IX Ch.5}}
\newblock (Springer-Verlag, Berlin Heidelberg, 2007).

\bibitem{baron;jpcs00}
Baron, A. Q. R. \emph{et al.} An X-ray scattering beamline for studying dynamics.
\newblock{\emph{J. Phys. Chem. Solids}} \textbf{61}, 461-465 (2000).

\bibitem{said;jsr11}
Said, A. H., Sinn, H. \& Divan, R. New developments in fabrication of high-energy-resolution analyzers for inelastic X-ray spectroscopy
\newblock {\emph{J. Synchrotron Radiat.}} \textbf{18}, 492-496 (2011).

\bibitem{kohn;hfi00}
Kohn, V. G. \& Chumakov, A. I. DOS: Evaluation of phonon density of states from nuclear resonant inelastic absorption.
\newblock{\emph{Hyperfine Interact.}} \textbf{125}, 205-221 (2000).

\bibitem{bosak;prb05}
Bosak, A. \& Krisch M. Phonon density of states probed by inelastic x-ray scattering.
\newblock{\emph{Phys. Rev. B}} \textbf{72}, 224305 (2005).

\bibitem{masad;prb07}
Masadeh, A. S. \emph{et al.} Quantitative size-dependent structure and strain determination of CdSe nanoparticles using atomic pair distribution function analysis.
\newblock{\emph{Phys. Rev. B}} \textbf{76}, 115413 (2007).

\bibitem{yang;pccp13}
Yang, X. \emph{et al.} Confirmation of disordered structure of ultrasmall CdSe nanoparticles from X-ray atomic pair distribution function analysis.
\newblock{\emph{Phys. Chem. Chem. Phys.}} \textbf{15}, 8480-8486 (2013).

\bibitem{han;jpcc12}
Han, P. \& Bester, G. Insights about the surface of colloidal nanoclusters from their vibrational and thermodynamic properties.
\newblock {\emph{J. Phys. Chem. C.}} \textbf{116(19)}, 10790-10795 (2012).

\bibitem{vozny;jpcc16}
Voznyy, O., Mokkath, J. H., Jain, A.,  Sargent, E. H. \& Schwingenschl\"ogl, U. Computational study of magic-size CdSe clusters with complementary passivation by carboxylic and amine ligands.
\newblock {\emph{J. Phys. Chem. C.}} \textbf{120}, 10015-10019 (2016).

\bibitem{toell;jsr11}
Toellner, T. S., Alatas, A. \& Said, A. H. Six-reflection meV-monochromator for synchrotron radiation.
\newblock {\emph{J. Synchrotron Radiat.}} \textbf{18}, 605-611 (2011).

\bibitem{chupa;jac03}
Chupas, P. J. \emph{et al.} Rapid-acquisition pair distribution function (RA-PDF) analysis.
\newblock {\emph{J. Appl. Crystallogr.}} \textbf{36}, 1342-1347 (2003).

\bibitem{hamme;hpr96}
Hammersley, A. P., Svenson, S. O., Hanfland, M. \& Hauserman, D. Two-dimensional detector software: from real detector to idealised image or two-theta scan.
\newblock {\emph{High Pressure Res.}} \textbf{14}, 235-248 (1996).

\bibitem{juhas;jac13}
Juh\'as, P., Davis, T., Farrow, C. L. \& Billinge, S. J. L. PDFgetX3: a rapid and highly automatable program for processing powder diffraction data into total scattering pair distribution functions.
\newblock {\emph{J. Appl. Crystallogr.}} \textbf{46}, 560-566 (2013).

\bibitem{egami;b;utbp13}
Egami, T. \& Billinge, S. J. L.
\newblock {\emph{Underneath the Bragg Peaks: Structural Analysis of Complex Materials}}
\newblock (Elsevier, Amsterdam, 2013).

\bibitem{bloch;prb94}
Bl\"ochl, P. E. Projector augmented-wave method.
\newblock{\emph{Phys. Rev. B}} \textbf{50}, 17953 (1994).

\bibitem{kress;prb96}
Kresse, G. \& Furthm\"uller J. Efficient iterative schemes for \emph{ab initio} total-energy calculations using a plane-wave basis set.
\newblock{\emph{Phys. Rev. B}} \textbf{54}, 11169 (1996).

\bibitem{deng;prb12}
Deng, H., Li, S., Li, J. \& Wei, S. Effect of hydrogen passivation on the electronic structure of ionic semiconductor nanostructures.
\newblock{\emph{Phys. Rev. B}} \textbf{85}, 195328 (2012).

\bibitem{baron;rmp01}
Baroni, S., de Gironcoli, S., Dal Corso, A. \& Giannozzi, P. Phonons and related crystal properties from density-functional perturbation theory.
\newblock{\emph{Rev. Mod. Phys.}} \textbf{73}, 515 (2001).

\bibitem{togo;prb08}
Togo, A. \& Tanaka, I. First principles phonon calculations in materials science.
\newblock{\emph{Scr. Mater.}} \textbf{108}, 1-5 (2015).

\end{thebibliography}
\end{document}